\documentclass[aps,pre,twocolumn,showpacs,groupedaddress,showkeys,amsmath,amssymb]{revtex4}
\usepackage{color}
\usepackage{epsfig}
\usepackage{graphicx}
\begin{document}
\title{On twinning in smectic crystals}
\author{V.I.\,Marchenko\\
Kapitza Institute for Physical Problems, RAS, Moscow, 119334
Russia}
\begin{abstract}
It is shown that mechanical twinning in smectic crystals is
possible. The structure of the boundary of twins for a small
disorientation of crystallites is determined. The periodic twin
structure, which should appear at the tension of the smectic
layer, is proposed.
\end{abstract}
\maketitle

The energy of small deformations of a smectic crystal is given by
the expression [1]
\begin{equation}\label{F} {\cal E}=\int\frac{A}{2}\left\{\left(\partial_zu
-\frac{(\partial_\alpha u)^2}{2}\right)^2+\lambda^2(\Delta_\perp
u)^2\right\}dV,
\end{equation}
where $u$ -- is the displacement of layers  along $z$ axis (in the
initial homogeneous undeformed state of the smectic crystal,
layers lie in the $xy$ plane, $A$ -- is the elastic modulus,
$\lambda$ -- is the length parameter, $\partial_\alpha$ -- is the
gradient vector in the $xy$ plane, and
${\Delta_\perp=\partial^2_\alpha.}$

According to Eq.\,(1), the undeformed state turned by small angle
${\theta\ll1}$ in the $xz$ plane (in this case,
${\partial_xu=\theta}$) corresponds to the derivative
${\partial_zu=\theta^2/2}.$ Let us consider the boundary between
the states ${\partial_xu=\pm\theta}$ ($x\rightarrow\pm\infty$)
that lies in the $yz$~plane. The quantity $\partial_zu$ is
unchanged inside the boundary. The variation equilibrium equation
in the problem under consideration reduces to the form
\begin{equation}\label{F2}
\lambda^2f'''+\frac{\theta^2}{2}f'-\frac{3}{2}f^2f'=0,\end{equation}
where ${f=\partial_xu}.$ The solution of this equation has the
form ${f=\theta\cdot\tanh(\theta x/2\lambda).}$ The energy of the
unit area of this boundary is given by the formula
\begin{equation}\label{Etwins}
\sigma=2A\lambda\theta^3/3.
\end{equation}

The twin structure of smectic crystals must be observed under the
conditions of Helfrich instability at strains noticeably larger
than the critical value (see the problem in [1,\,Sect.\,45]: the
smectic layer of thickness~$L$ bounded by solid walls parallel to
the smectic layer is extended along the $z$~axis). At very small
tensions ${\delta L>\delta L_c=2\pi\lambda},$ when $\delta L$ is
about the smectic period, the homogeneous state becomes unstable
with respect to the appearance of a periodic structure in the $xy$
plane with wave-number ${k_c=\sqrt{\pi/\lambda L}}.$ At a much
larger strain ${\delta L\gg\delta L_c},$ a twin structure as that
schematically shown in the figure should appear.

\centerline{\includegraphics[width=0.95\columnwidth]{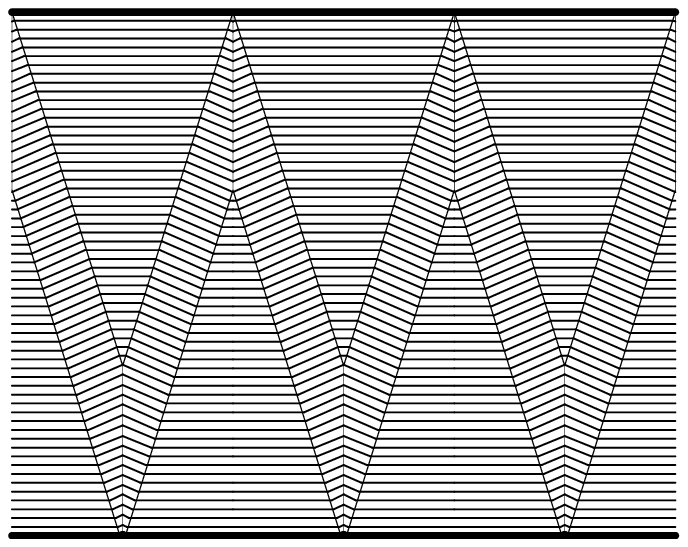}}

The parameters of this structure are determined by minimizing the
total energy of twin boundaries (${\theta=\varepsilon},$ at the
vertical boundaries and ${\theta=\varepsilon/2}$ at the boundaries
of the triangular regions). According to geometric consideration,
the angle at the vertex of a triangle is equal to $\varepsilon,$
the height $H$ of the triangles is related to the structure period
$d$ as ${\tan(\varepsilon/2)=d/2H},$ and the quantity $\delta L$
is related to the parameter $\varepsilon$ as
\[\delta L=(L-H)\left(\frac{1}{\cos\varepsilon}-1\right).\]

The energy density of the structure proposed above is given by the
expression
\begin{equation}\label{ee}
\frac{1}{Ld}\left\{2\frac{L-H}{\cos\varepsilon}\sigma(\varepsilon)+
4\frac{H}{\cos(\varepsilon/2)}\sigma(\varepsilon/2)\right\}.
\end{equation}

In view of the indicated geometric relationships, at a given
tension $\delta L/L$ energy (4) is a function of one parameter
$\varepsilon.$  When ${\delta L\ll L},$ angle $\varepsilon$ is
small. In this case, with the use of result (3), the minimum of
energy (4) is found to correspond to ${\varepsilon=\sqrt{6\delta
L/L}}.$ In this case, ${H=2L/3}$ and
\[d=2\sqrt{\frac{2L\delta L}{3}}.\]

I am grateful to E.I. Kats for stimulating discussions.

1. L.D. Landau and E.M. Lifshitz, Theory of Elasticity, Pergamon,
NY (1986)

Translated by R. Tyapaev

JETP Lett. {\bf116}(4), 587 (2007)
\end{document}